\documentclass[aps, prl, reprint]{revtex4-2}

\usepackage{graphicx}
\usepackage{color}
\usepackage{amssymb}
\usepackage{afterpage}
\usepackage{amsmath}
\usepackage{tabularx}
\usepackage{bm}
\usepackage{hyperref}
\usepackage{ltxtable}
\usepackage[caption=false]{subfig}
\usepackage{natbib} 
\usepackage{longtable}
\usepackage{float}
\usepackage{IEEEtrantools}
\newcommand{\be}{\begin{equation}}
\newcommand{\ee}{\end{equation}} 
\newcommand{\bes}{\begin{equation*}}
\newcommand{\ees}{\end{equation*}} 
\newcommand{\bea}{\begin{eqnarray}}
\newcommand{\eea}{\end{eqnarray}}

\usepackage[usenames,dvipsnames]{xcolor}
\hypersetup{
colorlinks,
citecolor=blue,
filecolor=blue,
linkcolor=blue,
urlcolor=blue}

\begin{document}
\title{Bistability in the sunspot cycle}
 
\author{Sumit Vashishtha}
\email{sumit.vashishtha@inria.fr}

\affiliation{Institut National de Recherche en Sciences et Technologies du Numérique (INRIA), Lille, France}

\author{Katepalli R. Sreenivasan}
\affiliation{Courant Institute of Mathematical Sciences, New York University, New York, USA}

\begin{abstract}{A direct dynamical test of the sunspot cycle is carried out to indicate that a
stochastically forced nonlinear oscillator characterizes its dynamics. The sunspot series is then decomposed into its eigen time-delay coordinates. The relevant analysis reveals that the sunspot series exhibits bistability, with the possibility of modeling the solar cycle as a stochastically and periodically forced bistable oscillator, accounting for poloidal and toroidal modes of the solar magnetic field. Such a representation enables us to conjecture stochastic resonance as the key mechanism in amplifying the planetary influence of Jupiter on the sun, and that extreme events, due to turbulent convection noise inside the sun, dictate crucial phases of the sunspot cycle, such as the Maunder minimum. }\end{abstract}

\maketitle

\begin{figure}
\centering
\includegraphics[width=7cm,height=10cm,keepaspectratio]{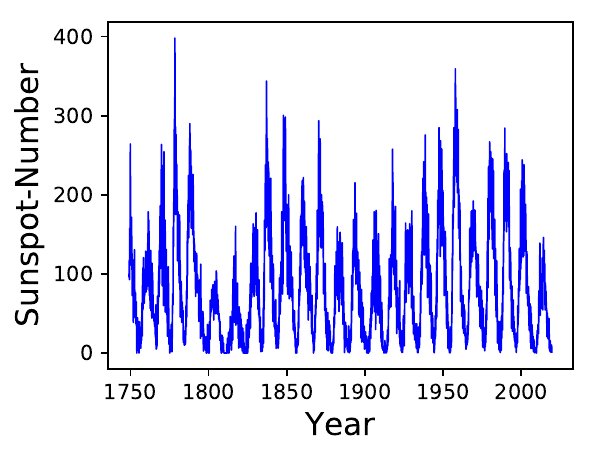}
\caption{Monthly-mean sunspot numbers. The dynamics causing the observed irregularity of amplitude with dominant periodicity in the time series remains unknown.}
\label{fig:sunspots} 
\end{figure}

 \textit{Introduction}.---Solar cycle prediction, such as forecasting the amplitude and/or the epoch of an upcoming  maximum, is of great importance for several reasons connected to space weather and possibly also to earth's climate ~\cite{kodera2002dynamical}.  However, such predictions have been quite inconclusive owing to inherent fluctuations in the period and amplitudes of each epoch of the solar cycle (Fig.~\ref{fig:sunspots}). Even though the global aspects of the solar cycle are explained by the dynamo theory~\cite{priest2012solar}, the nature of the irregularities displayed by the sunspot time series is still being debated, and detailed understanding of its dynamics is far from complete.

Some past work~\cite{knobloch1996new,mundt1991chaos} has claimed evidence for the origin of the sunspot cycle in deterministic chaos, based on estimations of correlation dimension, Lyapunov exponents, and an increase of a prediction error with a prediction horizon. However, these dimension-based algorithms have been found to be unreliable~\cite{theiler1986spurious,dammig1993estimation} when applied to relatively short experimental data, and properties consistent with stochastic processes (colored noises, etc) such as autocorrelations, can lead to spurious convergence of dimensional estimates. Moreover, the increase of a prediction error with an increasing prediction horizon is not a property exclusive for chaos. This behavior can also be observed in systems having a non-chaotic deterministic skeleton driven by a stochastic/noise component~\cite{gao1999can}. The aim of this letter is to characterize the role of noise and chaos in the sunspot cycle. Our point of departure is the so-called direct dynamical test~\cite{gao1999can, gao1994direct, gao2002noise} applied to sunspot time series.

\begin{figure}
\subfloat[\label{sfig:ddt_lorenz}]{%
	\includegraphics[width=0.65\linewidth]{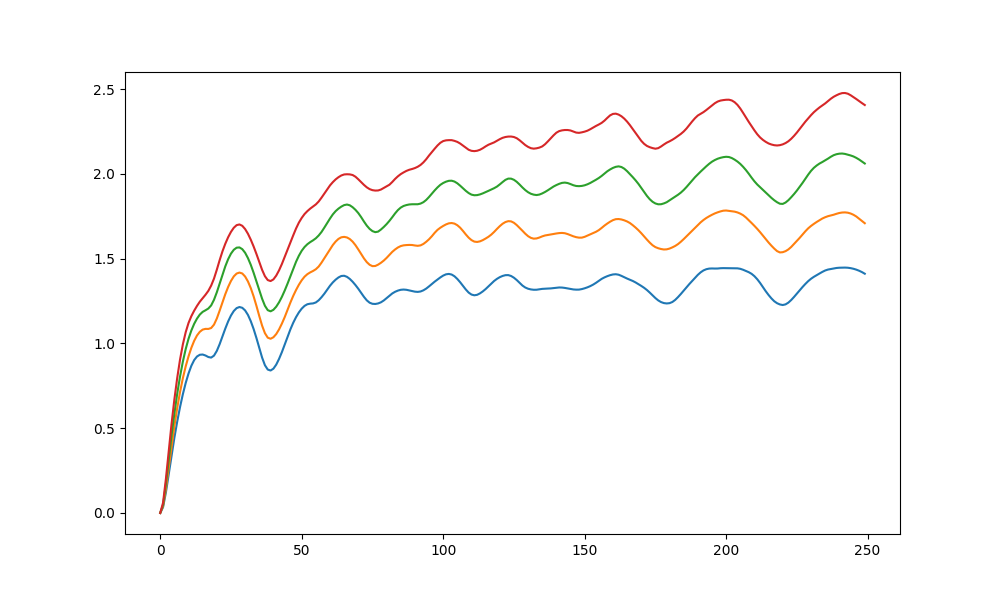}
}\\
\subfloat[\label{sfig:ddt_uniform_noise}]{%
	\includegraphics[width=0.65\linewidth]{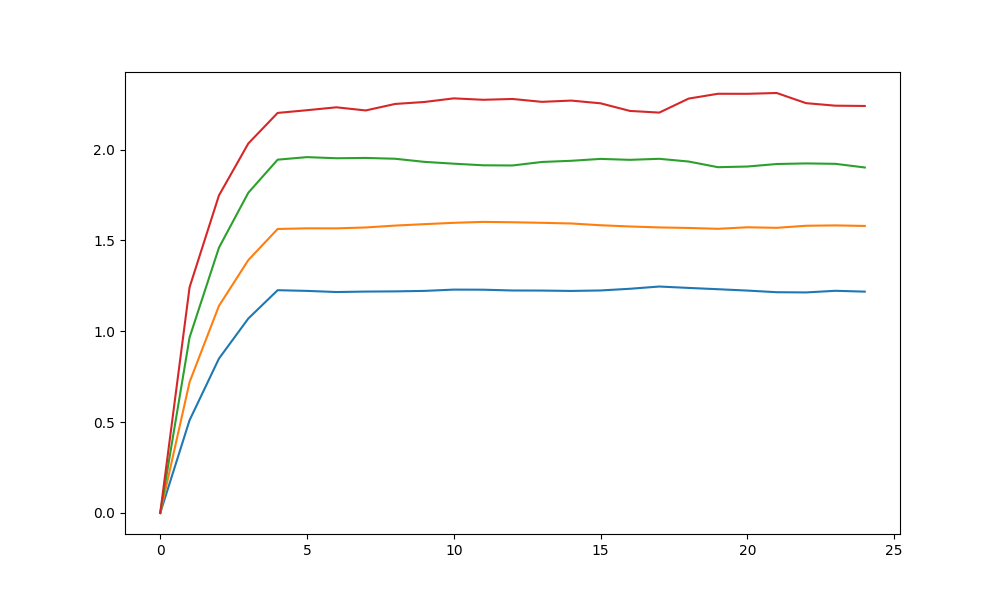}%
}
\caption{Divergence exponents, $\lambda(t)$, for (a) time series for a chaotic solution of the Lorenz system; for (b) random noise time series. Note that the linearly increasing portion of the plots overlap for Lorenz system exhibiting deterministic chaos in (a), whereas the lines break away from each other in (b).}
\label{fig:validate_ddt} 
\end{figure}

\textit{Direct Dynamical Test}.---It has been challenging to differentiate between noise and low-dimensional chaos. Reference \cite{gao1994direct} developed an effective test for distinguishing one from the other. We carried out this test for the sunspot series in order to infer the underlying dynamics. The application of the algorithm can be summarized as follows: From the  monthly-mean sunspot time series $\{x(i)\}$, we first construct vectors $\{X_i\}$ by the time delay embedding technique~\cite{packard1980geometry}: $X_{i} = [x(i),x(i+L),...,x(i+(m-1)L)]$, with $m$ as the embedding dimension and $L$ as the delay time. Utilizing the findings in~\cite{mundt1991chaos} in the context of the best values for delay time for the attractor reconstruction from the sunspot cycle, we
choose $L$ to be 10 in our computations. A value of 5 was used for $m$. We then compute
the time-dependent exponent, $\lambda(t)$, as

\be
\lambda(t) = \left \langle ln\left(\frac{\|X_{i+t}-X_{j+t}\|}{\|X_{i}-X_{j}\|}\right)\right\rangle,
\ee
with $r \leq \|X_i - X_j\| \leq r+\Delta r$, where $r$ and $\Delta r$ are
prescribed small distances. The angle brackets denote
ensemble averages of all possible pairs of $X_i$ and $X_j$. The
integer $t$, called the evolution time, corresponds to time
$t*dt$. Note that, geometrically, $(r, r + \Delta r)$ defines a shell, capturing the notion of scale. For clean chaotic systems, the $\lambda(t)$ curves first increase linearly with $t$ until some predictable time scale, $t_p$, is reached, and flatten~\cite{gao2002noise} thereafter. The linearly increasing parts of the $\lambda(t)$ curves corresponding to different shells collapse together to form an envelope for such clean systems. For noisy systems, the linearly increasing part of the $\lambda(t)$ curves, corresponding to small shells, break away from the envelope. The stronger the noise, the more conspicuously the $\lambda(t)$ curves break away from the envelope. Only if the noise is weak enough to allow the linearly increasing parts of
the $\lambda(t)$ curves, corresponding to some finite scale shells, to collapse together, can one say that the dynamics is chaotic.
This property forms a direct dynamical test for deterministic chaos~\cite{gao1994direct}. To illustrate this, we present a comparison of $\lambda(t)$ curves for the chaotic Lorenz system and random noise in Fig.~\ref{fig:validate_ddt}. Note that the linearly increasing parts of the curve collapse on each other for the Lorenz-system, but break apart for random noise. 

Note that $\lambda(t)$ gives a qualitative picture of the dynamics. Some quantitative aspects of the underlying dynamics can be assessed by the logarithmic displacement
\be
D(t) = ln\left\langle\left({\|X_{i+t}-X_{j+t}\|}\right)\right\rangle = \lambda(t) + ln\left\langle\left({\|X_{i}-X_{j}\|}\right)\right\rangle.
\ee
We comment subsequently on the behavior of $D(t)$. 

\begin{figure}
\subfloat[\label{sfig:diverge}]{%
	\includegraphics[width=0.65\linewidth]{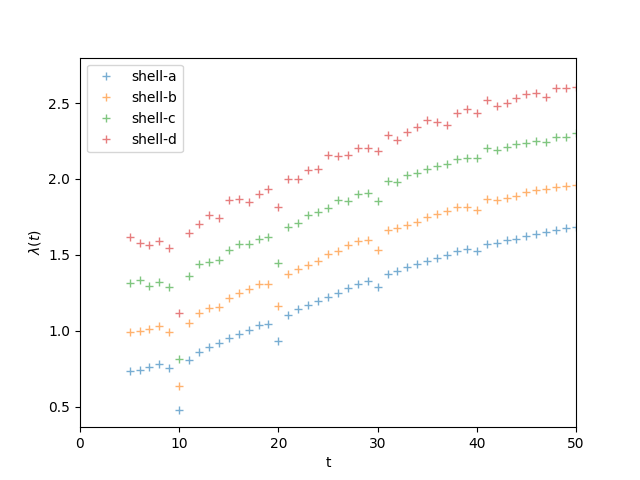}
}\\
\subfloat[\label{sfig:log_disp}]{%
	\includegraphics[width=0.65\linewidth]{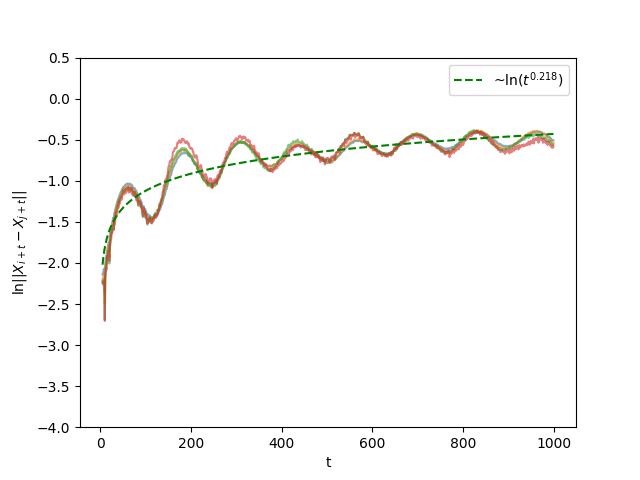}%
}
\caption{(a) Divergence exponents, $\lambda(t)$, for the sunspot time-series in the linear regime. Results for shells $(2^{-i/2}, 2^{-(i+1)/2})$, with $i = 7,8,9,10$ are shown. Other shells have similar results. (b) shows the logarithmic displacement curves for the sunspot time series. Note the sub-diffusive scaling for the curves. }
\label{fig:snn_ddt} 
\end{figure}

Now we carry out the same analysis for the monthly-mean sunspot time series. Figure~\ref{fig:snn_ddt}(a) shows the $\lambda (t)$ exponents (in the linearly increasing region) for four different shells, and Fig.~\ref{fig:snn_ddt}(b) exhibits the $D(t)$ curves for the same shells. The two conclusions that can be drawn are: (a) Sunspot series are not necessarily chaotic, and the dynamics is greatly influenced by noise. This can be inferred from Fig.~\ref{fig:snn_ddt}(a) as follows: If the time series were to exhibit deterministic chaos, all the plots should have collapsed on each other for the linearly increasing portion. This is not the case, thereby weakening the case for deterministic chaos, suggesting an important role of noise in the sunspots dynamics. (b) Sunspot series exhibits anomalous diffusion~\cite{dubkov2008levy}. This is exhibited in Fig~\ref{fig:snn_ddt}(b), wherein the temporal evolution of logarithmic displacement is plotted. Note that this displacement scales (on the average) as $t^{\alpha}$, where $\alpha=0.218$, thus implying sub-diffusion. This sub-diffusive scaling is observed in systems such as stochastically-driven nonlinear oscillators~\cite{gao1999noise}. Overall, the direct dynamical test shows that any possibility of chaos is overpowered by the effect of noise in the sunspot series, and in a strict sense, the sunspot series do not exhibit deterministic chaos; rather a stochastically driven nonlinear oscillator better describes the evolution of the sunspot time series. Observations supporting this argument have been made in the past \cite{paluvs1999sunspot, mininni2000stochastic} as well, but, as shown in ~\cite{zou2014complex}, these relaxation-oscillator models could not provide a complete description of the solar cycle dynamics. A way to characterize this stochastic oscillator is to decompose the sunspot series into its eigen time-delay coordinates, inspired by Koopman operator theory, as we shall describe next.

\textit{Eigen time-delay coordinates}.---Consider a dynamical system of the form
\be
\label{eq:Dynamical_system}
\frac{\rm{d}}{{{\rm d}t}}{\bf{x}}(t) = {\bf{f}}({\bf{x}}(t)),
\ee

the discretized form of which is given as

\be
\label{eq:Dynamical_system_discrete}
{{\bf{x}}_{k + 1}} = {\bf{F}}({{\bf{x}}_k}) = {{\bf{x}}_k} + {\int}_{k\Delta t}^{(k + 1)\Delta t} {\bf{f}}({\bf{x}}(\tau ))\rm{d}\tau.
\ee
Here $x(t)\in \mathbb{R}^{n}$ is the state of the system at time $t$ and $f$ represents the dynamic constraints that define the equations of motion. There are two major perspectives for analyzing such system: (a) The traditional geometric perspective of dynamical systems, which describes the topological organization of trajectories of $\mathbf{x}$, mediated by fixed points, periodic orbits, and attractors of the dynamics f; and (b) the evolution of measurements of the state, $y=g(x)$, using the perspective provided by Koopman ~\cite{koopman1931hamiltonian, mezic2013analysis,mezic2005spectral}. The latter analysis relies on the existence of a linear operator $\cal K$ for the dynamical system in Eq.~\ref{eq:Dynamical_system_discrete}, given by
\be
\label{eq:Koopman_domain}
{\cal K}g \buildrel \Delta \over = g \circ {\bf{F}}\quad \Rightarrow \quad {\cal K}g({{\bf{x}}_k}) = g({{\bf{x}}_{k + 1}}).
\ee

The Koopman operator $\cal K$ induces a linear system on the space of all measurement functions g, and transforms the finite-dimensional nonlinear dynamics in Eq.~\ref{eq:Dynamical_system} to an infinite-dimensional linear dynamics in Eq.~\ref{eq:Koopman_domain}, providing a global linear representation valid far away from fixed points and periodic orbits. Obtaining a finite-dimensional approximation of the Koopman operator is challenging, and a Koopman-invariant measurement system is the key for such a realization. Eigen time-delay coordiantes have been shown to approximate a Koopman-invariant measurement system, and used to construct the best fit linear models for various dynamical systems in the past~\cite{kutz2016dynamic,brunton2017chaos}, using simple linear regression. In this instance, these eigen time-delay
coordinates may be obtained from monthly-mean sunspot time series
$\{x(t_1), x(t_2), x(t_3), \ldots\}$, by taking a singular value decomposition of the Hankel matrix $\mathrm{H}$
\begin{IEEEeqnarray}{rCl}
\label{eq:SVD}
\mathrm{H} =
\begin{bmatrix}
  x(t_1) & x(t_2) & \ldots &x(t_{p})  \\
  x(t_2)  & x(t_3)  &  \ldots & x(t_{p+1}) \\
  \vdots  &\vdots  & \ddots & \vdots \\ 
 x(t_q) & x(t_{q+1}) & \ldots & x(t_{m})
\end{bmatrix}
= \mathrm{U\Sigma V^{*}.}
\end{IEEEeqnarray}
An important hyperparameter above is the number of delays $q$, chosen such that the delay duration $D = (q - 1) \Delta t$ is large enough to capture a sufficient duration of the oscillation, where $\Delta t$ is the sampling period. As a rule of thumb~\cite{champion2019discovery}, $q$ should be chosen such that $D = T$, where $T$ is the time period of the signal. We choose $q$ such that $D$ is slightly greater than 11 years, which is the average period for the sunspot series. Equation \ref{eq:SVD} yields a hierarchical decomposition of the matrix H into eigen time-series given by the columns of U and V. These columns are ordered by their ability to express the variance in the columns and rows of H, respectively. The relative importance of each of these columns is expressed by the eigenvalue diagonal matrix $\Sigma$ containing the singular-values, ${\sigma_{i}^2}$.

\begin{figure}
\subfloat[\label{sfig:Eigenvalues}]{%
	\includegraphics[width=0.8\linewidth]{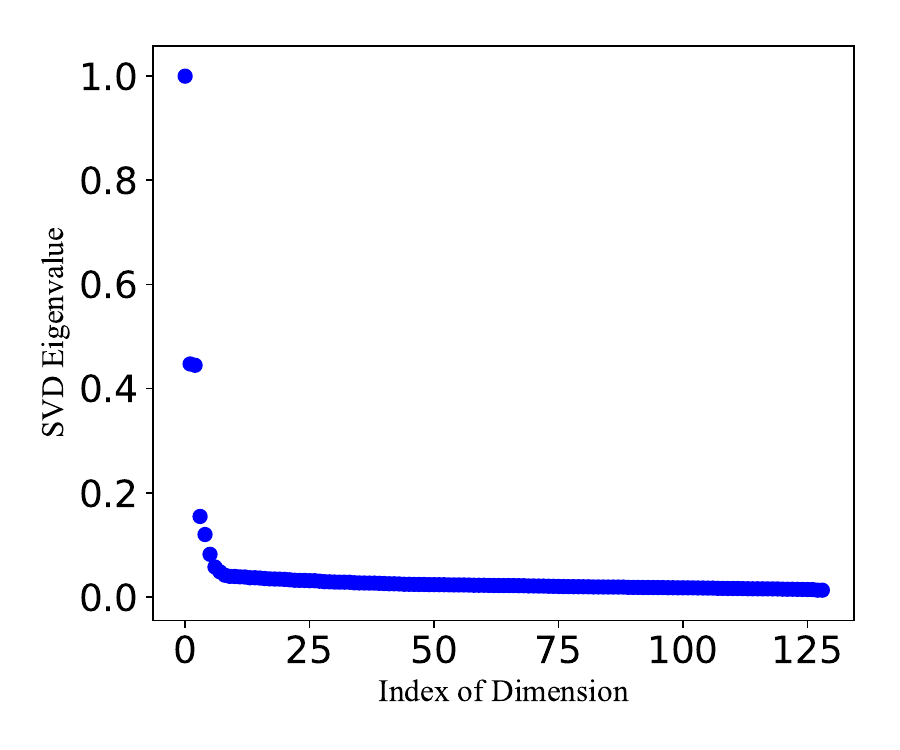} }
\\
\subfloat[\label{sfig:Eigen-time-series}]{%
	\includegraphics[width=0.8\linewidth]{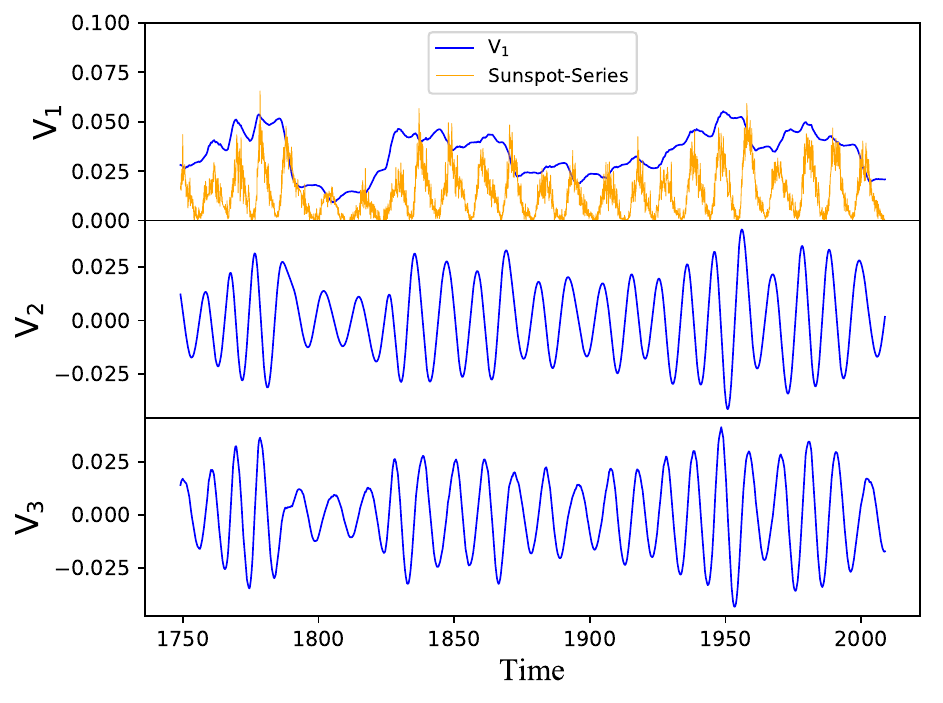}%
}
\caption{(a) Distribution of the coordinate $\mathrm{V_{1}}$. Note the bimodal nature of the distribution. This implies bistability and the possibility of the underlying dynamics being governed by a forced potential-well system; (b) The first three dominant delay coordinates ($\mathrm{V_{1}}, \mathrm{V_{2}}, \mathrm{V_{3}}$). Note that $\mathrm{V_{1}}$ is the amplitude envelope of the sunspot series, whereas $\mathrm{V_{2}}$ and $\mathrm{V_{3}}$ are periodic signals with their amplitudes modulated by $\mathrm{V_{1}}$.}
\label{fig:SVD} 
\end{figure}

\begin{figure}
\subfloat[\label{sfig:V1-Hist}]{%
	\includegraphics[width=0.8\linewidth]{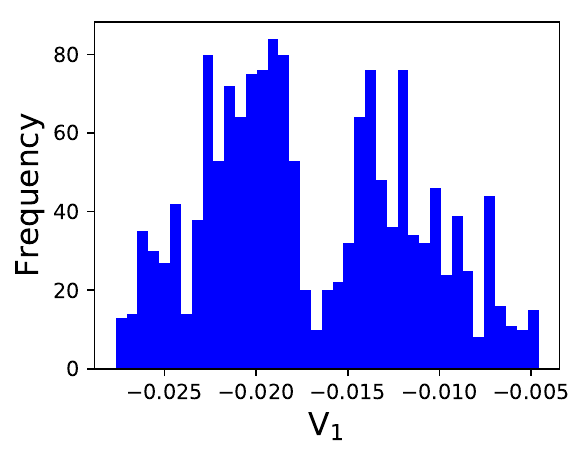}
}\\
\subfloat[\label{sfig:Sunspots-Spectrum}]{%
	\includegraphics[width=0.8\linewidth]{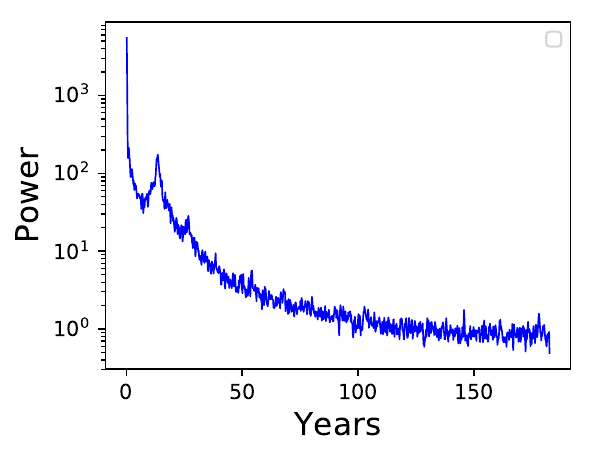}}
\caption{(a) Distribution of the coordinate $\mathrm{V_{1}}$. Note its bimodal nature. This implies bistability and the possibility of the underlying dynamics being governed by a forced quartic potential well system; (b) Spectra of the sunspot series.  }
\label{fig:Stochastic-Res} 
\end{figure}
Figure~\ref{sfig:Eigenvalues} shows the eigenvalue spectrum, which makes it clear that the dynamics underlying the solar cycle is low-dimensional given that only the first two or three eigenvalues are significant. The columns of $\mathrm{V}$ provide a time series of the magnitude of each of the columns of $\mathrm{U}\Sigma$ in the data. The time series for the columns of V corresponding to dominant eigenvalues, or the leading delay coordinates, is shown in Fig.~\ref{sfig:Eigen-time-series}. The first mode $\mathrm{V_1}$ turns out closely to be an amplitude envelope of the original sunspot series. The other two modes, besides having some phase lag, appear close to each other. These modes are essentially periodic signals, with their amplitudes modulated by $\mathrm{V_1}$. Figure~\ref{fig:Stochastic-Res}(a) shows the distribution of different values in the time-series of $\mathrm{V_1}$. The bimodal nature of the histogram suggests bistability in the sunspot series. Figure~\ref{sfig:Sunspots-Spectrum} shows the spectrum for the sunspot cycle. Aside from the main period and its harmonics, the spectrum essentially exhibits a noisy background. 

These observations lead us to propose the following one-dimensional nonlinear oscillator model for the sunspot series:
\be
 \frac{dx}{dt} = -\frac{\partial U}{\partial x} + F(t) + \sum_{i}A_{i}\mathrm{sin}{\omega_{i}}t
 \label{Eq:Quartic-oscillator}.
 \ee
Here the first term on the right hand side is the restoring force, with $U = -\alpha x^{2} + \beta x^{4}$ as the quartic potential function corresponding to a bistable system. Along with a periodic external forcing, a random component $F(t)$ is also present.  

\begin{figure}[ht]
\subfloat[\label{sfig:V1-derivative-pdf}]{%
	\includegraphics[width=0.8\linewidth]{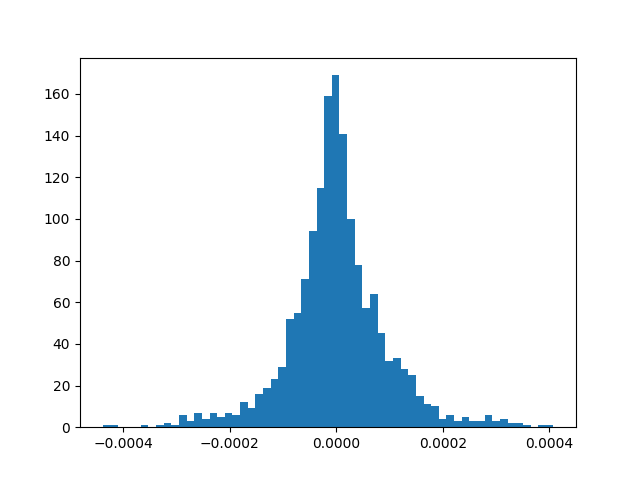}
}\\
\subfloat[\label{sfig:V1-derivative-pdf-probability-paper}]{%
\includegraphics[width=0.8\linewidth]{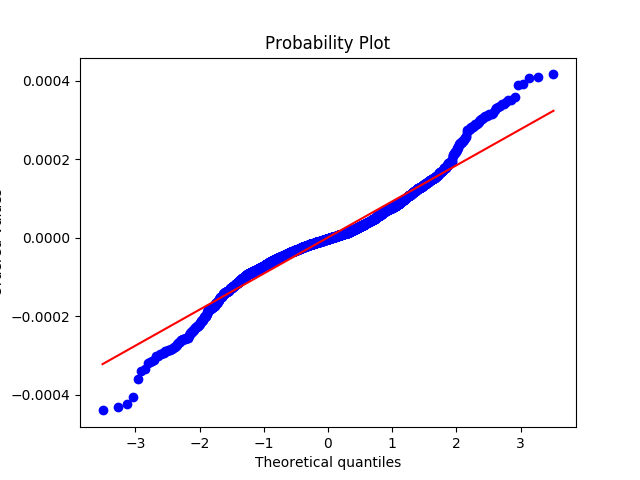}%
}
\caption{(a) Distribution of the time-derivative of the eigen delay-coordinate $\mathrm{V_{1}}$; (b) Same distribution on a probability scale}
\label{fig:V1-derivative-pdf} 
\end{figure}

The nature of this noise term (to first order) can be inferred by analyzing the time derivative of the eigen time-series $\mathrm{V_1}$ (Eq.~\ref{Eq:Quartic-oscillator} without periodic forcing). The pdf of the derivative of this delay coordinate is shown in Fig.~\ref{fig:V1-derivative-pdf}a. The same pdf is also shown on a scale where Gaussian distribution is a straight line (probability paper scale) in Fig.~\ref{fig:V1-derivative-pdf}b. The deviation from the straight line shows that the nature of noise term driving the bistable dynamics proposed above is non-Gaussian and heavy-tailed. The bistable dynamics underlying the sunspot series would account for the poloidal and toroidal components of the solar magnetic field. Finally, note that Eq.~\ref{Eq:Quartic-oscillator} can also exhibit stochastic resonance, wherein a sub-threshold periodic signal can be entrained in the dynamics because of the additive role of noise. The characteristic of Fig.~\ref{sfig:Sunspots-Spectrum}, showing a spectrum of noisy background, with peaks at driving frequencies and its harmonics, is characteristic for systems exhibiting stochastic resonance~\cite{wiesenfeld1995stochastic,benzi1981mechanism}. The ratio of the eigenvalues corresponding to stochastic and periodic components in the present case is around 2.5, highlighting the weak contribution from the periodic-component of the forcing, and dominating role of the noise term. Because of the limited amount of data available for the sunspot series, standard analysis for stochastic resonance, such as residence time distribution, does not yield any meaningful information. Even so, if stochastic resonance occurs in the above model for the sunspot cycle, it could serve as a mechanism for the amplification of weak planetary influences on the sun.
 
\textit{The possibility of weak planetary influence on the solar cycle}.---There is a striking similarity between the average revolution time period of the Jupiter (around 11.86 years) and the (noisy) periodicity of the sunspot cycle. Given this similarity, the possible role of the (very weak) planetary forcing by Jupiter in influencing the solar magnetic cycle cannot be ignored, and has been studied in the past~\cite{stefani2016synchronized,abreu2012there, callebaut2012influence, wilson2008does}. However, no clear physical mechanism for how such weak planetary forcing could make itself so dominantly evident in the solar cycle. One possible mechanism is stochastic-resonance~\cite{benzi1981mechanism}, wherein a very weak external periodic signal is entrained in the dynamics at some optimal level of noise inside the system. A system exhibiting stochastic resonance should typically be bistable, which is what we have reported in this paper. Thus, in light of that finding, stochastic resonance appears to be a plausible mechanism for the weak periodic planetary forcing from Jupiter to influence the solar-cycle. This effect can be naturally incorporated in models of the type in Eq.~\ref{Eq:Quartic-oscillator}. Note that the bistability reported in this paper would correspond to the poloidal and toroidal components of the solar magnetic field. In fact, many state-of-the-art models in solar-dynamo theory are based on transformations of the toroidal components to the poloidal components of the magnetic field, and the other way around. See ~\cite{silva1993bipolar, choudhuri1995solar, choudhuri2007predicting} for more details. Further, note that the noise in model~\ref{Eq:Quartic-oscillator} would correspond to the noise of turbulent convection in the sun.

\textit{Maunder minimum as rare events}.---We saw in Fig.~\ref{sfig:V1-derivative-pdf-probability-paper} that the noise-driven bistable dynamics causes a heavy-tailed sunspot cycle. Thus, rare events may play an important role in the evolution of the sunspot cycle. Consider, for instance, Maunder minimum~\cite{hoyt1996well, ribes1993solar}, a phase of grand minima in the sunspot cycle during 1645–1715, when the solar activity was strongly reduced. It has been established through an analysis of geological records that several Maunder minimum like periods have occurred in the more remote past. It may well turn out that rare events in the stochastic forcing, representing the extreme events in turbulent convection, drive such phases. For instance, an extreme event can confine the dynamics to the potential well corresponding to poloidal component of the magnetic field. This will result in significant reduction in the toroidal component (which directly corresponds to the number of sunspots (see ~\cite{choudhuri1995solar})) of the magnetic field, and hence in the number of sunspots observed. A similar argument can be made to explain the phase where the values of maxima were very high in the sunspot cycle. Such rare event driven dynamics have been shown to play an important role in the dynamics of climate~\cite{ditlevsen1999observation}, transition to turbulence in pipe flows~\cite{barkley2016theoretical}, and in aerodynamic bifurcations~\cite{gayout2020rare}, among others.

\textit{Conclusions}.---We have shown that the sunspot series exhibits bistability. First, a direct dynamical test~\cite{gao1994direct} of the sunspot series indicated that a forced nonlinear oscillator governs its dynamics. After that, we carried out an analysis of the dominant eigen time-delay coordinates of the sunspot series, from which we concluded that the oscillator is likely to be a one-dimensional bistable oscillator driven by heavy-tailed random forcing and weak periodic forcing. Such a stochastic bistable dynamical system representation of the sunspot series enabled us to conjecture stochastic resonance as the key mechanism in amplifying the planetary influence of Jupiter on the sun, and that rare events in the turbulent convection noise inside the sun could dictate crucial phases of the sunspot cycle, such as the Maunder minimum. Our findings strongly encourage modeling attempts of the solar cycle that incorporate the possibility of nonlinear effects such as stochastic resonance ~\cite{albert2021can}.

%\textit{Acknowledgements}


%\bibliography{chaos_bibliography}
%\bibliographystyle{apsrev4-1}

\end{document}